\documentclass{article}
\usepackage{ismir,amsmath,cite,url}
\usepackage{graphicx}
\usepackage{color}
\usepackage{booktabs} % For formal tables
\usepackage[flushleft]{threeparttable}
\usepackage{amssymb}

\title{Multi-label Music Genre Classification from audio, text, and images using Deep Features}
\oneauthor
 {Sergio Oramas$^1$, Oriol Nieto$^2$, Francesco Barbieri$^3$, Xavier Serra$^1$} {$^1$Music Technology Group, Universitat Pompeu Fabra \\ $^2$Pandora Media Inc. \\ $^3$TALN Group, Universitat Pompeu Fabra \\
 \{sergio.oramas, francesco.barbieri, xavier.serra\}@upf.edu, onieto@pandora.com}

\begin{document}

\maketitle
\begin{abstract}
Music genres allow to categorize musical items that share common characteristics. Although these categories are not mutually exclusive, most related research is traditionally focused on classifying tracks into a single class. Furthermore, these categories (e.g., Pop, Rock) tend to be too broad for certain applications. In this work we aim to expand this task by categorizing musical items into multiple and fine-grained labels, using three different data modalities: audio, text, and images. To this end we present \textit{MuMu}, a new dataset of more than 31k albums classified into 250 genre classes. For every album we have collected the cover image, text reviews, and audio tracks. Additionally, we propose an approach for multi-label genre classification based on the combination of feature embeddings learned with state-of-the-art deep learning methodologies. Experiments show major differences between modalities, which not only introduce new baselines for multi-label genre classification, but also suggest that combining them yields improved results.

\end{abstract}

\section{Introduction}\label{sec:introduction}
Music genres are useful labels to classify musical items into broader categories that share similar musical, regional, or temporal characteristics. Dealing with large collections of music poses numerous challenges when retrieving and classifying information \cite{casey2008content}. 
Music streaming services tend to offer catalogs of tens of millions of tracks, for which tasks such as music classification are of utmost importance.
Music genre classification is a widely studied problem in the Music Information Research (MIR) community \cite{sturm2012survey}. 
However, almost all related work is concentrated in multi-class classification of music items into broad genres (e.g., Pop, Rock), assigning a single label per item. This is problematic since there may be hundreds of more specific music genres \cite{pachet2000taxonomy}, and these may not be necessarily mutually exclusive (i.e., a song could be Pop, and at the same time have elements from Deep House and a Reggae grove). 
In this work we aim to advance the field of music classification by framing it as multi-label genre classification of fine-grained genres.

% One of the most relevant aspects of deep learning approaches is that they are able to learn new data representations from raw data, avoiding the traditional dependency on the feature extraction step. 
% This aspect makes deep learning methodologies more domain independent than traditional machine learning approaches \cite{bengio2013representation}. 
% Therefore, methodologies used in one domain can be applied to other domains with slight modifications, opening up vast possibilities for multimodal approaches in MIR.

% Music classification of a handful of broad genres may have been an interesting computational problem for a while, but now it is time to move forward to a more real world problem such as multi-label genre classification of very specific genres.

To this end, we present \emph{MuMu}, a new large-scale multimodal dataset for multi-label music genre classification. \emph{MuMu} contains information of roughly 31k albums classified into one or more 250 genre classes. For every album we analyze the cover image, text reviews, and audio tracks, with a total number of approximately 147k audio tracks and 447k album reviews. 
Furthermore, we exploit this dataset with a novel deep learning approach to learn multiple genre labels for every album using different data modalities (i.e., audio, text, and image). 
In addition, we combine these modalities to study how the different combinations behave.

Results show how feature learning using deep neural networks substantially surpasses traditional approaches based on handcrafted features, reducing the gap between text-based and audio-based classification \cite{oramas2016exploring}.
Moreover, an extensive comparative of different deep learning architectures for audio classification is provided, including the usage of a dimensionality reduction approach that yields improved results. 
% One of the main contribution of this work is the usage of a dimensionality reduction approach of sparse target labels. We demonstrate how this approach improves classification in terms of accuracy and catalog coverage. 
Finally, we show how the late fusion of feature vectors learned from different modalities achieves better scores than each of them individually.

%The paper is structured as follows...

\section{Related Work}\label{sec:related}

%Music genre classification is a widely studied task within the Music Information Research (MIR) community. However, 
Most published music genre classification approaches rely on audio sources \cite{sturm2012survey,bogdanov2016cross}. 
Traditional techniques typically use handcrafted audio features, such as Mel Frequency Cepstral Coeﬃcients (MFCCs) \cite{logan2000mel}, as input of a machine learning classifier (e.g., SVM) \cite{tzanetakis2002musical,seyerlehner2010using}.
More recent deep learning approaches take advantage of visual representations of the audio signal in form of spectrograms.
These visual representations are used as input to Convolutional Neural Networks (CNNs) \cite{dieleman2011audio,dieleman2014end,pons2016experimenting,Choi2016,choi2016convolutional}, following approaches similar to those used for image classification.

Text-based approaches have also been explored for this task. 
For instance, in \cite{hu2005mining,oramas2016exploring} album customer reviews are used as input for the classification, whereas in \cite{mayer2008rhyme,choi2014song} song lyrics are employed.
By contrast, there are a limited number of papers dealing with image-based genre classification \cite{libeks2011you}.
Most multimodal approaches for this task found in the literature combine audio and song lyrics as text \cite{laurier2008multimodal,neumayer2007integration}. 
Moreover, the combination of audio and video has also been explored \cite{schindler2015audio}. 
However, the authors are not aware of published multimodal approaches for music genre classification that involve deep learning. 

Multi-label classification is a widely studied problem \cite{tsoumakas2006multi,jain2016extreme}. 
Despite the scarcity in terms of approaches for multi-label classification of music genres \cite{Sanden2011,wang2009tag}, there is a long tradition in MIR for tag classification, which is a multi-label problem \cite{Choi2016,wang2009tag}.

\section{Multimodal Dataset}\label{sec:mumu}

To the best of our knowledge, there are no publicly available large-scale datasets that encompass audio, images, text, and multi-label annotations. % with enough data to be suitable for approaches that tend to require large amounts of data (e.g., deep learning). 
Therefore, we present \emph{MuMu}, a new Multimodal Music dataset with multi-label genre annotations that combines information from the Amazon Reviews dataset \cite{mcauley2015image} and the Million Song Dataset (MSD) \cite{Bertin-Mahieux2011}. 
The former contains millions of album customer reviews and album metadata gathered from Amazon.com. 
The latter is a collection of metadata and precomputed audio features for a million songs. 

To map the information from both datasets we use MusicBrainz\footnote{https://musicbrainz.org/}. 
For every album in the Amazon dataset, we query MusicBrainz with the album title and artist name to find the best possible match. Matching is performed using the same methodology described in \cite{Oramas2015b}, following a pair-wise entity resolution approach based on string similarity. Following this approach, we were able to map 60\% of the Amazon dataset.
%60\% of albums were mapped in the dataset. 
For all the matched albums, we obtain the MusicBrainz recording ids of their songs. 
With these, we use an available mapping from MSD to MusicBrainz\footnote{http://labs.acousticbrainz.org/million-song-dataset-echonest-archive} to obtain the subset of recordings present in the MSD. 
From the mapped recordings, we only keep those associated with a unique album.
This process yields the final set of 147,295 songs, which belong to 31,471 albums.

The song features provided by the MSD are not generally suitable for deep learning \cite{Oord2013}, so we instead use in our experiments audio previews between 15 and 30 seconds retrieved from \texttt{7digital.com}.
For the mapped set of albums, there are 447,583 customer reviews in the Amazon Dataset. 
In addition, the Amazon Dataset provides further information about each album, such as genre annotations, average rating, selling rank, similar products, cover image url, etc. 
We employ the provided image url to gather the cover art of all selected albums. 
%All of these data %(i.e., audio, text, and images) 
The mapping between the three datasets (Amazon, MusicBrainz, and MSD), genre annotations, data splits, text reviews, and links to images are released as the \emph{MuMu} dataset\footnote{https://www.upf.edu/web/mtg/mumu}. Images and audio files can not be released due to copyright issues.

\subsection{Genre Labels}\label{sec:taxonomy}

Amazon has its own hierarchical taxonomy of music genres, which is up to four levels in depth.
In the first level there are 27 genres, and almost 500 genres overall. 
In our dataset, we keep the 250 genres that satisfy the condition of having been annotated in at least 12 albums. %instances in the training set (TODO: what are these sets? What's their split \%?), one in the validation set and one in the test set. 
Every album in Amazon is annotated with one or more genres from different levels of the taxonomy. 
The Amazon Dataset contains complete information about the specific branch from the taxonomy used to classify each album. For instance, an album annotated as Traditional Pop comes with the complete branch information \textit{Pop / Oldies / Traditional Pop}. 
To exploit either the taxonomic and the co-occurrence information, we provide every item with the labels of all their branches. For example, an album classified as \textit{Jazz / Vocal Jazz} and \textit{Pop / Vocal Pop} is annotated in \emph{MuMu} with the four labels: Jazz, Vocal Jazz, Pop, and Vocal Pop. There are in average 5.97 labels for each song (3.13 standard deviation).

\begin{table}[!htp]
\centering
\scriptsize
\caption{Top-10 most and least represented genres}
\label{tbl:genres}
\begin{tabular}{lr|lr}
\toprule
Genre & \% of albums & Genre & \% of albums \\
\midrule
Pop & 84.38 & Tributes & 0.10 \\
Rock & 55.29 & Harmonica Blues & 0.10 \\
Alternative Rock & 27.69 & Concertos & 0.10 \\
World Music & 19.31 & Bass & 0.06 \\
Jazz & 14.73 & European Jazz & 0.06 \\
Dance \& Electronic & 12.23 & Piano Blues & 0.06 \\
Metal & 11.50 & Norway & 0.06 \\
Indie \& Lo-Fi & 10.45 & Slide Guitar & 0.06 \\
R\&B & 10.10& East Coast Blues & 0.06 \\
Folk & 9.69 & Girl Groups & 0.06 \\
\bottomrule
\end{tabular}
\end{table}

The labels in the dataset are highly unbalanced, following a distribution which might align well with those found in real world scenarios. 
%but this is the common case in real world scenarios. 
In Table~\ref{tbl:genres} we see the top 10 most and least represented genres and the percentage of albums annotated with each label.
The unbalanced character of the genre annotations poses an interesting challenge for music classification that we also aim to exploit. 
Among the multiple possibilities that this dataset may offer to the MIR community, we focus our work on the multi-label classification problem, described next.

% In addition, the availability of different data modalities allows us to study the problem from different perspectives. 

%(TODO: Acabar la sección diciendo que, aunque unbalanced, este set podría ser muy útil en el campo de Music Classification, pues no sólo hay diversas fuentes de información a explotar, si no que se podría enfocar como un multi-label problem, described next.)

\section{Multi-label Classification}\label{sec:multilabel}

%(TODO: Esta sección debería definir formalmente el problema de Multi-label Music Classification, y no hablar de la approach que presentamos en el paper. Creo que la approach debería estar descrita en la siguiente sección. Aquí puedes hablar de la evaluación y tal vez enfocar la descripción de PMI para otros domains besides music. Se podría dividir así:
%4.1. Problem definition
%4.2. Methods in other domains
%4.3. Evaluation Metrics).

In multi-label classification, multiple target labels may be assigned to each classifiable instance. %Framing this as a music genre classification problem, we can formally define it as follows: 
More formally: given a set of $n$ labels $L = \{l_1,l_2,\ldots,l_n\}$, and a set of $m$ items $I = \{i_1,i_2,\ldots,i_m\}$, we aim to model a function $f$ able to associate a set of $c$ labels to every item in $I$, where $c \in [1, n]$ varies for every item. %This problem may be seen as multiple single-label binary classification problems, one per label.
%Traditional approaches for multi-label classification tend to separate it into multiple single-label binary classification problems, in a one-vs-all way.

Deep learning approaches are well-suited for this problem, as these architectures allow to have multiple outputs in their final layer.
%To this end, we train a deep neural network able to predict the different labels. 
The usual architecture for large multi-label classification using deep learning ends with a logistic regression layer with sigmoid activations evaluated with the cross-entropy loss, where target labels are encoded as high-dimensional sparse binary vectors \cite{szegedy2016rethinking}. 
This method, which we refer as \textsc{logistic}, implies the assumption that the classes are statistically independent (which is not the case in music genres).
%, which is not the case in music genres. Co-occurrence and taxonomic relations between genres avoid this independence, and this relations are not exploited following this approach.
%Therefore, 

A more recent approach \cite{Chollet2016}, relies on matrix factorization to reduce the dimensionality of the target labels. 
This method makes use of the interrelation between labels, embedding the high-dimensional sparse labels onto lower-dimensional vectors.
In this case, the target of the network is a dense lower-dimensional vector which can be learned using the cosine proximity loss, as these vectors tend to be $l2$-normalized. 
We denote this technique as \textsc{cosine}, and we provide a more formal definition next.

\subsection{Labels Factorization}\label{sec:factorization}

%Music genres are typically organized in a taxonomy. In these taxonomies, when a musical item belongs to a genre (e.g. Power-Pop), it also belongs to its parents (e.g. Pop). 
%The typical approach for large multi-label classification problems using deep learning have a final logistic regression layer with a sigmoid cross-entropy loss, with target labels encoded as high-dimensional sparse binary vectors. This method implies the assumption that the classes are statistically independent, which is not the case. Co-occurrence and taxonomic relations between labels avoid this independence, and this is not exploited following this approach.

%We apply a method for dimensionality reduction of the label space that takes into account the interrelation between labels in the dataset. 
%To embed our high-dimensional sparse labels onto lower-dimensional vectors we applied a method for dimensionality reduction 

% Among the different methods in the literature for dimensionality reduction, we focus on the approach presented in \cite{Levy2011}, and further applied to the multi-label classification problem in \cite{Chollet}. 
Let $M$ be the binary matrix of items $I$ and labels $L$ where $m_{ij} = 1$ if $i_i$ is annotated with label $l_j$ and $m_{ij} = 0$ otherwise. Using $M$, we calculate the matrix $X$ of Positive Pointwise Mutual Information (PPMI) for the set of labels $L$. Given $L_i$ as the set of items annotated with label $l_i$, the PPMI between two labels is defined as:

\begin{equation}
X(l_i,l_j) =
max\left(0,\log{\frac{P(L_i,L_j)
  }{
    P(L_i)P(L_j)
  }}\right)
\end{equation}

where $P(L_i,L_j) = |L_i \cap L_j| / |I|$ and $P(L_i) = |L_i| / |I|$.

The PPMI matrix $X$ is then factorized using Singular Value Decomposition (SVD) such that $X \approx U \Sigma V$, where $U$ and $V$ are unitary matrices, and $\Sigma$ is a diagonal matrix of singular values. Let $\Sigma_d$ be the diagonal matrix formed from the top $d$ singular values, and let $U_d$ be the matrix produced by selecting the corresponding columns from $U$, the matrix $C_d = U_d \cdot \sqrt{\Sigma_d}$ contains the label factors of $d$ dimensions. Finally, we obtain the matrix of item factors $F_d$ as $F_d = C_d \cdot M^T$. Further information on this technique may be found in \cite{levy2014neural}.

Factors present in matrices $C_d$ and $F_d$ are embedded in the same space. Thus, a distance metric such as cosine distance can be used to obtain distance measures between items and labels. Similar labels are grouped in the space, and at the same time, items with similar sets of labels are near each other. These properties can be exploited in the label prediction problem.

\subsection{Evaluation Metrics}\label{sec:metrics}

The evaluation of multi-label classification is not necessarily straightforward. 
Evaluation measures vary according to the output of the system. 
In this work we are interested in measures that deal with probabilistic outputs, instead of binary. 
The Receiver Operating Characteristic (ROC) curve is a graphical plot that illustrates the performance of a binary classifier system as its discrimination threshold is varied. 
Thus, the area under the ROC curve (AUC) is often taken as an evaluation measure to compare such systems. 
We selected this metric to compare the performance of the different approaches as it has been widely used for genre and tag classification problems \cite{Choi2016,dieleman2014end}. 
%Although other measures such as the area under the precision-recall curve may be more adequate for unbalanced classes \cite{saito2015precision}, we use AUC-ROC to evaluate our approaches, as this measure has been widely used for genre and tag classification problems within the MIR community \cite{Choi2016,dieleman2014end}. 
%Therefore, we selected this metric to compare the performance of the different approaches.

The output of a multi-label classifier is a label-item matrix. 
Thus, it can be evaluated either from the labels or the items perspective. 
We can measure how accurate the classification is for every label, or how well the labels are ranked for every item. 
In this work, the former point of view is evaluated with the AUC measure, which is computed for every label and then averaged. 
We are interested in classification models that strengthen the diversity of label assignments. 
As the taxonomy is composed of broad genres which are over-represented in the dataset (see Table~\ref{tbl:genres}), and more specific subgenres (e.g., Vocal Jazz, Britpop), we want to measure whether the classifier is focusing only on over-represented genres, or on more fine-grained ones.
To this end, catalog coverage (also known as aggregated diversity) is an evaluation measure used in the extreme multi-label classification \cite{jain2016extreme} and the recommender systems \cite{oramas2016sound} communities. %It is called either aggregated diversity or catalog coverage. %We want to measure how diverse is assignation of labels to items.
Coverage@k measures the percentage of normalized unique labels present in the top $k$ predictions made by an algorithm across all test items. Values of $k = 1, 3, 5$ are typically employed in multi-label classification.

%It can be evaluated from different perspectives, from the labels side or from the items side. We can measure how accurate is the classification in every class, or how well are the labels ranked for every item. 
%(TODO: describe the most important evaluation metrics)
%Different approaches have been used for evaluation. The area under the ROC curve (AUC-ROC) has been widely used in classification problems, and also to some tag classification problems within the MIR community \cite{Keun, Dieleman}. Therefore, we selected this metric to compare the performance of the different approaches.

%However, according to \cite{paper roc }, ROC plots could be misleading when applied in imbalanced classification scenarios like ours. They propose instead to use the precision-recall curve (PRC). They argue that these kind of plots provide a more real insight of the actual performance of the system, as they evaluate the fraction of true positives among positive predictions. We used then in our experiments the area under the precision-recall curve (AUC-PRC), also called Average Precision score (AP). We present the average of the AUC-PRC across all the classes.

%The multi-label classification problem can be seen as a recommendation problem, where we want to recommend labels to items. Therefore, we can also use measures coming from the recommender systems community such as catalog coverage. 

\section{Album Genre Classification}\label{sec:classification}

In this section we exploit the multimodal nature of the \emph{MuMu} dataset to address the multi-label classification task.
More specifically, and since each modality on this set (i.e., cover image, text reviews, and audio tracks) is associated with a music album, our task focuses on album classification. %classifying albums.%this type of musical items.

\subsection{Audio-based Approach}\label{sec:audio}

A music album is composed by a series of audio tracks, each of which may be associated with different genres. %, but these generally tend to be closely related to each other. % (TODO: cite? no se me ocurre). 
% In the \emph{mumu} dataset we have genre annotations at the album level. 
In order to learn the album genre from a set of audio tracks we split the problem into three steps: (1) track feature vectors are learned while trying to predict the genre labels of the album from every track in a deep neural network. (2) Track vectors of each album are averaged to obtain album feature vectors.
%compute the centroid of all feature vectors of tracks from the same album, given rise to an album feature vector. 
(3) Album genres are predicted from the album feature vectors in a shallow network where the input layer is directly connected to the output layer.

%(TODO: This paragraph should go in the lit review section) There are several approaches in the literature for genre classification from audio using deep learning. 
%We want to compare some of the main contributions of these works. 
It is common in MIR to make use of CNNs to learn higher-level features from spectrograms. 
These representations are typically contained in $\mathbb{R}^{\mathcal{F} \times N}$ matrices with $\mathcal{F}$ frequency bins and $N$ time frames.
In this work we compute 96 frequency bin, log-compressed constant-Q transforms (CQT) \cite{Schorkhuber2010} for all the tracks in our dataset using \texttt{librosa} \cite{Mcfee2015} with the following parameters: audio sampling rate at 22050 Hz, hop length of 1024 samples, Hann analysis window, and 12 bins per octave.
In addition, log-amplitude scaling is applied to the CQT spectrograms.
Following a similar approach to \cite{Oord2013}, we address the variability of the length $N$ across songs by sampling one 15-seconds long \emph{patch} from each track, resulting in the fixed-size input to the CNN.

To learn the genre labels we design a CNN with four convolutional layers and experiment with different number of filters, filter sizes, and output configurations (see Section~\ref{sec:audioexp}).
%% Removed to reduce space
%These networks are trained using mini batches of 32 items, randomly sampled from the training data to compute the gradient, and Adam \cite{KingmaB14} is the optimizer used to train the models, with the default suggested learning parameters.

%(TODO: This paragraph seems like it should go in the experiments section. Or at least after describing the input to the net and what the net is for)

%The audio classification task in Section \ref{} was performed on the track level. To obtain album feature vectors from track vectors we...

%We take the best logistic regression approach \textsc{class-low-4x70} for compatibility with other modalities because the ResNet used to learn the visual features has this output configuration.

\subsection{Text-based Approach}
\label{sec:text}
In the presented dataset, each album has a variable number of customer reviews. 
We use an approach similar to \cite{hu2005mining,oramas2016exploring} for genre classification from text, where all reviews from the same album are aggregated into a single text. 
The aggregated result is truncated at 1000 characters, thus balancing the amount of text per album, as more popular artists tend to have a higher number of reviews. %(TODO: What balance? Why 1000?). 
Then we apply a Vector Space Model approach (VSM) with tf-idf weighting \cite{Zobel1998} to create a feature vector for each album. 
Although word embeddings \cite{Mikolov2013} with CNNs are state-of-the-art in many text classification tasks \cite{Kim2014}, a traditional VSM approach is used instead, as it seems to perform better when dealing with large texts \cite{Oramas2017}.
The vocabulary size is limited to 10k as it was a good balance of network complexity and accuracy.% (adding more words increased the complexity of the network but did not significantly improved the results).
%<----- TODO: ANADIDO ESTO PERO SE PUEDE QUITAR

Furthermore, a second approach is proposed based on the addition of semantic information, similarly to the method described in \cite{oramas2016exploring}. 
To semantically enrich the album texts, we adopted Babelfy, a state-of-the-art tool for entity linking \cite{Moroetal2014}, a task to associate, for a given textual fragment candidate, the most suitable entry in a reference KB. 
Babelfy maps words from a given text to Wikipedia\footnote{http://wikipedia.org}. 
In Wikipedia, categories are used to organize resources. %, and they help users to group articles of the same subject. 
We take all the Wikipedia categories of entities identified by Babelfy in each document and add them at the end of the text as new words. 
Then a VSM with tf-idf weighting is applied to the semantically enriched texts, where the vocabulary is also limited to 10k terms. 
Note that either words or categories may be part of this vocabulary. 
%Babelfy \cite{}, a state-of-the-art Entity Linking system, is applied to the review texts and the Wikipedia categories of every entity identified by Babelfy are added at the end of the text as new words. Then a VSM with tf-idf weighting is applied to the semantically enriched texts, where the vocabulary is also limited to 10k terms. Note that either words or categories may be part of this vocabulary. 

From this representation, a feed forward network with two dense layers of 2048 neurons and a Rectified Linear Unit (ReLU) after each layer is trained to predict the genre labels in both \textsc{logistic} and \textsc{cosine} configurations. 

%In addition, we compared the performance of the deep neural network against a one-vs-all VSM classifier.

\subsection{Image-based Approach}
\label{sec:resnet}
Every album in the dataset has an associated cover art image. To perform music genre classification from these images, we use Deep Residual Networks (ResNets) \cite{he2016deep}.
They are the state-of-the-art in various image classification tasks like Imagnet \cite{ILSVRC15} and Microsoft COCO \cite{lin2014microsoft}. 
ResNet is a common feed-forward CNN with \emph{residual learning}, which consists on bypassing two or more convolution layers. %The residual connections limit the underfitting problem originated when using a high number of layers. This way, the number of layer can be increased, creating very deep CNNs (more than 100 layers).
We employ a slightly modified version of the original ResNet\footnote{https://github.com/facebook/fb.resnet.torch/}: the scaling and aspect ratio augmentation are obtained from \cite{szegedy2015going}, the photometric distortions from \cite{howard2013some}, and weight decay is applied to all weights and biases. %(i.e., not focusing on convolutional layers only).
The network we use is composed of 101 layers (ResNet-101), initialized with pretrained parameters learned on ImageNet.
This is our starting point to finetune the network on the genre classification task.
Our ResNet implementation has a logistic regression final layer with sigmoid activations and uses the binary cross entropy loss.
%% Removed to gain space
%The network is trained on the genre classification task with mini batches of 50 samples for 90 epochs, a learning rate of 0.0001, and with Adam\cite{KingmaB14} as optimizer.

\subsection{Multimodal approach}\label{sec:multimodal}

We aim to combine all of these different types of data into a single model.
There are several works claiming that learning data representations from different modalities simultaneously outperforms systems that learn them separately \cite{ngiam2011multimodal,dorfer2016towards}. However, recent work in multimodal learning with audio and text in the context of music recommendation \cite{Oramas2017} reflects the contrary. We have observed that deep networks are able to find an optimal minimum very fast from text data. However, the complexity of the audio signal can significantly slow down the training process. Simultaneous learning may under-explore one of the modalities, as the stronger modality may dominate quickly. Thus, learning each modality separately warrants that the variability of the input data is fully represented in each of the feature vectors.

Therefore, from each modality network described above, we separately obtain an internal feature representation for every album after training them on the genre classification task.
Concretely, the input to the last fully connected layer of each network becomes feature vector for its respective modality.
Given a set of feature vectors, $l2$-regularization is applied on each of them. 
They are then concatenated into a single feature vector, which becomes the input to a simple Multi Layer Perceptron (MLP), where the input layer is directly connected to the output layer. 
The output layer may have either a \textsc{logistic} or a \textsc{cosine} configuration. %Similar to the audio and text networks, the output layer has 250 units (to match the number of classes aiming to be predicted) with sigmoid activation and binary cross entropy loss for the \textsf{logistic} configuration, and 50 units with linear activation and cosine proximity loss for the \textsf{cosine} configuration. We empirically set up 50\% of dropout after the input in the \textsf{cosine} setting, and no dropout in the \textsf{logistic} setting.

\begin{table*}[ht]
\centering
\begin{threeparttable}
\scriptsize
\caption{Results for Multi-label Music Genre Classification of Albums}
\label{tbl:results}
\begin{tabular}{lcccccccc}
Modality & Target & Settings                & Params         & Time    & AUC        & C@1           & C@3           & C@5           \\
\toprule
\textsc{Audio} & \textsc{logistic} & \textsc{timbre-mlp}                   &  0.01M              & 1s              & 0.792               & 0.04             & 0.14              & 0.22        \\
\textsc{Audio} & \textsc{logistic} & \textsc{low-3x3}           & 0.5M           & 390s          & 0.859          & 0.14          & 0.34          & 0.54          \\
\textsc{Audio} & \textsc{logistic} & \textsc{high-3x3}          & 16.5M          & 2280s         & 0.840          & 0.20          & 0.43          & 0.69          \\
\textsc{Audio} & \textsc{logistic} & \textsc{low-4x96}          & 0.2M           & 140s          & 0.851          & 0.14          & 0.32          & 0.48          \\
\textsc{Audio} & \textsc{logistic} & \textsc{high-4x96}         & 5M             & 260s          & 0.862          & 0.12          & 0.33          & 0.48          \\
\textsc{Audio} & \textsc{logistic} & \textsc{low-4x70} & 0.35M & 200s & 0.871 & 0.05 & 0.16 & 0.34 \\
\textsc{Audio} & \textsc{logistic} & \textsc{high-4x70}         & 7.5M           & 600s          & 0.849          & 0.08          & 0.23          & 0.38          \\
\textsc{Audio} & \textsc{cosine} & \textsc{low-3x3}            & 0.33M          & 400s          & 0.864          & 0.26          & 0.47          & 0.65          \\
\textsc{Audio} & \textsc{cosine} & \textsc{high-3x3}           & 15.5M          & 2200s         & 0.881          & 0.30          & 0.54          & 0.69          \\
\textsc{Audio} & \textsc{cosine} & \textsc{low-4x96}           & 0.15M          & 135s          & 0.860          & 0.19          & 0.40          & 0.52          \\
\textsc{Audio} & \textsc{cosine} & \textsc{high-4x96}          & 4M             & 250s          & 0.884          & 0.35          & 0.59          & 0.75          \\
\textsc{Audio} & \textsc{cosine} & \textsc{low-4x70}           & 0.3M           & 190s          & 0.868          & 0.26          & 0.51          & 0.68          \\
\textbf{\textsc{Audio (A)}} & \textbf{\textsc{cosine}} & \textbf{\textsc{high-4x70}} & \textbf{6.5M}  & \textbf{590s} & \textbf{0.888} & \textbf{0.35} & \textbf{0.60} & \textbf{0.74} \\
\midrule
%Text & logistic & one-vs-all SVM & & & & & & \\
\textsc{Text} & \textsc{logistic} & \textsc{VSM} & 25M & 11s & 0.905 & 0.08 & 0.20 & 0.37 \\
\textsc{Text} & \textsc{logistic} & \textsc{VSM+Sem} & 25M & 11s & 0.916 & 0.10 & 0.25 & 0.44 \\
\textsc{Text} & \textsc{cosine} & \textsc{VSM} & 25M & 11s & 0.901 & 0.53 & 0.44 & 0.90 \\
\textbf{\textsc{Text (T)}} & \textbf{\textsc{cosine}} & \textbf{\textsc{VSM+Sem}} & \textbf{25M} & \textbf{11s} & \textbf{0.917} & \textbf{0.42} & \textbf{0.70} & \textbf{0.85} \\
\midrule
\textsc{Image (I)} & \textsc{logistic} & \textsc{ResNet} & 1.7M & 4009s & 0.743 & 0.06 & 0.15 & 0.27 \\
%\textbf{Image} & \textbf{cosine} & \textbf{ResNet} & \textbf{1.7M} & \textbf{4009s} & \textbf{0.744} & \textbf{0.08} & \textbf{0.22} & \textbf{0.31} \\
\midrule
\textsc{A + T} & \textsc{logistic} & \textsc{mlp} & 1.5M & 2s & 0.923 & 0.10 & 0.40 & 0.64 \\
\textsc{A + I} & \textsc{logistic} & \textsc{mlp} & 1.5M & 2s & 0.900 & 0.10 & 0.38 & 0.66 \\
\textsc{T + I} & \textsc{logistic} & \textsc{mlp} & 1.5M & 2s & 0.921 & 0.10 & 0.37 & 0.63 \\
\textbf{\textsc{A + T + I}} & \textbf{\textsc{logistic}} & \textbf{\textsc{mlp}} & \textbf{2M} & \textbf{2s} & \textbf{0.936} & \textbf{0.11} & \textbf{0.39} & \textbf{0.66} \\
\textsc{A + T} & \textsc{cosine} & \textsc{mlp} & 0.3M & 2s & 0.930 & 0.43 & 0.74 & 0.86 \\
\textsc{A + I} & \textsc{cosine} & \textsc{mlp} & 0.3M & 2s & 0.896 & 0.32 & 0.57 & 0.76 \\
\textsc{T + I} & \textsc{cosine} & \textsc{mlp} & 0.3M & 2s & 0.919 & 0.43 & 0.74 & 0.85 \\
\textsc{A + T + I} & \textsc{cosine} & \textsc{mlp} & 0.4M & 2s & 0.931 & 0.42 & 0.72 & 0.86 \\
\bottomrule
\end{tabular}
    \begin{tablenotes}
      \small
      \item Number of network hyperparameters, epoch training time, AUC-ROC, and catalog coverage at $k = 1,3,5$ for different settings and modalities.
    \end{tablenotes}
   \end{threeparttable}
\end{table*}

\section{Experiments}\label{sec:experiments}

We apply the architectures defined in the previous section to the \emph{MuMu} dataset. 
The dataset is divided as follows: 80\% for training, 10\% for validation, and 10\% for test. We first evaluate every modality in isolation in the multi-label genre classification task. 
Then, from each modality, a deep feature vector is obtained for the best performing approach in terms of AUC. 
Finally, the three modality vectors are combined in a multimodal network. 
All results are reported in Table~\ref{tbl:results}. 
Performance of the classification is reported in terms of AUC score and Coverage@k with $k = 1, 3, 5$. 
The training speed per epoch and number of network hyperparameters are also reported. 
All source code and data splits used in our experiments are available on-line\footnote{https://github.com/sergiooramas/tartarus}.

\begin{figure}[!htp]
\centerline{
\includegraphics[width=0.9\columnwidth]{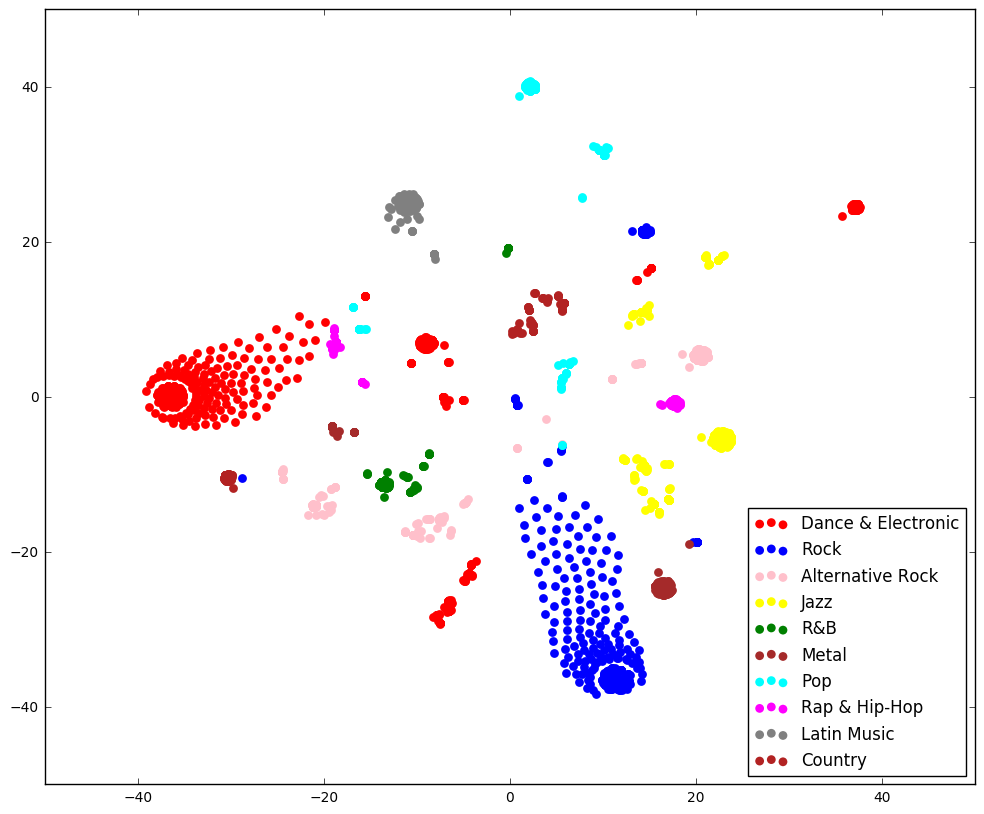}}
\caption{t-SNE of album factors.}
\label{fig:tsne}
\end{figure}

The matrix of album genre annotations of the training and validation sets is factorized using the approach described in Section~\ref{sec:factorization}, with a value of $d = 50$ dimensions.
From the set of album factors, those annotated with a single label from the top level of the taxonomy are plotted in Figure~\ref{fig:tsne} using t-SNE dimensionality reduction \cite{maaten2008visualizing}.
It can be seen how the different albums are properly clustered in the factor space according to their genre.

\subsection{Audio Classification}\label{sec:audioexp}

We explore three network design parameters: convolution filter size, number of filters per convolutional layer, and target layer. 
For the filter size we compare three approaches: square 3x3 filters as in \cite{Choi2016}, a filter of 4x96 that convolves only in time \cite{Oord2013}, and a musically motivated filter of 4x70, which is able to slightly convolve in the frequency domain \cite{pons2016experimenting}. 
To study the width of the convolutional layers we try with two different settings: \textsc{high} with 256, 512, 1024, and 1024 in each layer respectively, and \textsc{low} with 64, 128, 128, 64 filters. %We empirically set these two configurations. 
Max-pooling is applied after each convolutional layer.
Finally, we use the two different network targets defined in Section~\ref{sec:multilabel}, \textsc{logistic} and \textsc{cosine}. %Moreover, we studied how the use of Dropout regularization \cite{} affect both target configurations.
We empirically observed that dropout regularization only helps in the \textsc{high} plus \textsc{cosine} configurations. Therefore we applied dropout with a factor of 0.5 to these configurations, and no dropout to the others. %The dataset is standarized with zero mean and unit variance.

Apart from these configurations, a baseline approach is added. This approach consists in a traditional audio-based approach for genre classification based on the audio descriptors present in the MSD \cite{Bertin-Mahieux2011}.
More specifically, for each song we aggregate four different statistics of the 12 timbre coefficient matrices: mean, max, variance, and $l2$-norm.
The obtained 48 dimensional feature vectors are fed into a feed forward network as the one described in Section~\ref{sec:multimodal} with \textsc{logistic} output.
This approach is denoted as \textsc{timbre-mlp}.
%Second, we added a state-of-the-art multi-label genre classification approach based on the combination of convolutional and recurrent neural networks \cite{Keun2} whose source code is available\footnote{\url{https://github.com/keunwoochoi/music-auto_tagging-keras}}. We computed the mel-spectrograms with the provided code and trained the model using the exact same architecture, just changing the size of the output layer match the number of genre labels.

%Results of the different approaches are reported in Table~\ref{tbl:audio}. 

The results show that CNNs applied over audio spectrograms clearly outperform traditional approaches based on handcrafted features. 
We observe that the \textsc{timbre-mlp} approach achieves 0.792 of AUC, contrasting with the 0.888 from the best CNN approach.
We note that the \textsc{logistic} configuration obtains better results when using a lower number of filters per convolution (\textsc{low}). Configurations with fewer filters have less parameters to optimize, and their training processes are faster. 
On the other hand, in \textsc{cosine} configurations we observe that the use of a higher number of filters tends to achieve better performance. 
It seems that the fine-grained regression of the factors benefits from wider convolutions.
Moreover, we observe that 3x3 square filter settings have lower performance, need more time to train, and have a higher number of parameters to optimize.
By contrast, networks using time convolutions only (\textsc{4x96}) have a lower number of parameters, are faster to train, and achieve comparable performance. 
Furthermore, networks that slightly convolve across the frequency bins (\textsc{4x70}) achieve better results with only a slightly higher number of parameters and training time. 
Finally, we observe that the \textsc{cosine} regression approach achieves better AUC scores in most configurations, and also their results are more diverse in terms of catalog coverage.

%As described in Section~\ref{}, album feature vectors are created by averaging the feature vectors of the different tracks of an album. In Table~\ref{} we evaluate this approach by comparing different strategies for feature aggregation. We compare these strategies in the \textsc{logistic/low-4x70} and \textsc{cosine/high-4x70} respectively. We observe in the results that the average outperform the maximum and the random selection of a track feature vector.

%(TODO: Tabla)

\subsection{Text Classification}\label{sec:textexp}

For text classification, we obtain two feature vectors as described in Section~\ref{sec:text}: one built from the texts \textsc{VSM}, and another built from the semantically enriched texts \textsc{VSM+Sem}. 
Both feature vectors are trained in the multi-label genre classification task using the two output configurations \textsc{logistic} and \textsc{cosine}.

Results show that the semantic enrichment of texts clearly yields better results in terms of AUC and diversity.
Furthermore, we observe that the \textsc{cosine} configuration slightly outperforms \textsc{logistic} in terms of AUC, and greatly in terms of catalog coverage. 
The text-based results are overall slightly superior to the audio-based ones. 

We also studied the information gain of words in the different genres. We observed that genre labels present in the texts have important information gain values. However, it is remarkable that \textit{band} is a very informative word for Rock, \textit{song} for Pop, and \textit{dope}, \textit{rhymes}, and \textit{beats} are discriminative features for Rap albums. Place names have also important weights, as \textit{Jamaica} for Reggae, \textit{Nashvile} for Country, or \textit{Chicago} for Blues.%\footnote{The complete list of words is available on-line at https://www.upf.edu/en/web/mtg/mumu}.

\subsection{Image Classification}\label{sec:imageexp}

\begin{figure}
\centering
\includegraphics[height=6.5cm,keepaspectratio]{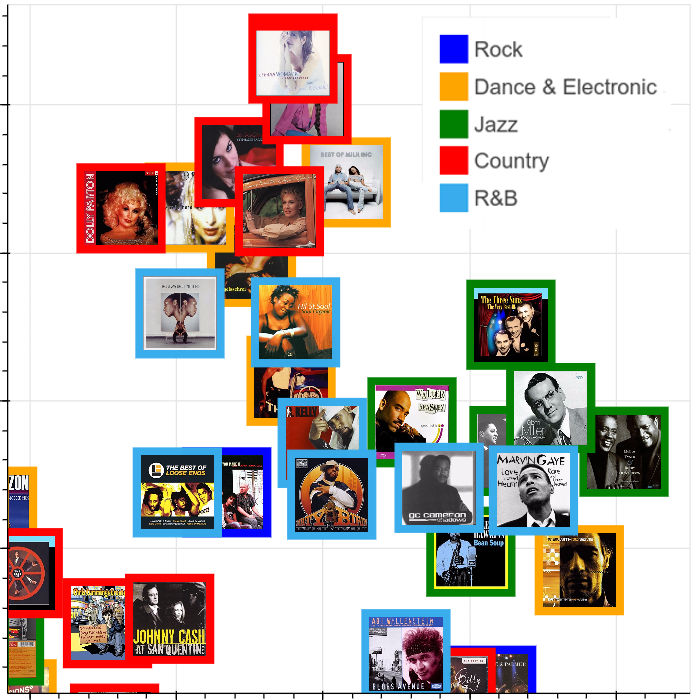} \\ 
\caption{Particular of the t-SNE of randomly selected image vectors from five of the most frequent genres.}
\label{fig:tsne_visual}
\end{figure}

%(TODO: We need better consistency: these parameters should go in the previous section, as with the ones of audio. Alternative, you can move those from section 5.1 to 6.1).
Results show that genre classification from images has lower performance in terms of AUC and catalog coverage compared to the other modalities. Due to the use of an already pre-trained network with a logistic output (ImageNet \cite{ILSVRC15}) as initialization of the network, it is not straightforward to apply the \textsc{cosine} configuration. Therefore, we only report results for the \textsc{logistic} configuration.

In Figure~\ref{fig:tsne_visual} a set of cover images of five of the most frequent genres in the dataset is shown using t-SNE over the obtained image feature vectors. 
In the left top corner the ResNet recognizes women faces on the foreground, which seems to be common in Country albums (red).
%Also the R\&B genre appears to be generally well clustered, since black men that the network sucessfully recognizes tend to appear on the cover. 
The jazz albums (green) on the right are all clustered together probably thanks to the uniform type of clothing worn by the people of their covers. 
% In general, the visual content can give us relevant information about the singer, and this can be very informative. %(like for the African-American in R\&B, or also in Rap \& Hip Hop where the singer often wear a backwards hats).
Therefore, the visual style of the cover seems to be informative when recognizing the album genre.
For instance, many classical music albums include an instrument in the cover, and Dance \& Electronics covers are often abstract images with bright colors, rarely including human faces.

\subsection{Multimodal Classification}\label{sec:multiexp}

From the best performing approaches in terms of AUC of each modality (i.e., \textsc{Audio} / \textsc{cosine} / \textsc{high-4x70}, \textsc{Text} / \textsc{cosine} / \textsc{VSM+Sem} and \textsc{Image} / \textsc{logistic} / \textsc{ResNet}), a feature vector is obtained as described in Section~\ref{sec:multimodal}. 
Then, these three feature vectors are aggregated in all possible combinations, and genre labels are predicted using the MLP network described in Section~\ref{sec:multimodal}.
Both output configurations \textsc{logistic} and \textsc{cosine} are used in the learning phase, and dropout of 0.7 is applied in the \textsc{cosine} configuration.
%Although a feature vector can be extracted from a network trained with \textsc{cosine} configuration (e.g. \textsc{Audio} / \textsc{cosine} / \textsc{high-4x70}), when this vector is used in a multimodal approach, the multimodal network can be either trained with \textsc{logistic} or \textsc{cosine} configurations. The same happens with vectors trained with \textsc{logistic} loss.

Results suggest that the combination of modalities outperforms single modality approaches. 
As image features are learned using a \textsc{logistic} configuration, they seem to improve multimodal approaches with \textsc{logistic} configuration only. 
Multimodal approaches that include text features tend to improve the results. %, even if the difference in performance between the different combinations is not pronounced.
Nevertheless, the best approaches are those that exploit the three modalities of \emph{MuMu}. \textsc{Cosine} approaches have similar AUC than \textsc{logistic} approaches but a much better catalog coverage, thanks to the spatial properties of the factor space. 

%Album factors tend to be near their associated genre factors in the latent space, and similar genres are grouped together. Therefore, when an album factor is predicted by the network, the nearest genres in the space achieve the highest probability score. Although popular genre factors may have some hub properties in the latent space \cite{radovanovic2010hubs}, they can not be nearest neighbor of the majority of album factors. However, in the \textsc{logistic} configuration, it is easy for the network to prioritize one specific output that is very popular in the dataset (e.g. Pop or Rock), because this warranties a reduction of the crossentropy loss.

\section{Conclusions}\label{sec:conclusions}

An approach for multi-label music genre classification using deep learning architectures has been proposed. 
The approach was applied to audio, text, image data, and their combination. 
For its assessment, \emph{MuMu}, a new multimodal music dataset with over 31k albums and 135k songs has been gathered. 
% This dataset encompasses multimodal information of about . Results show how the different modalities behave in the task. 
We showed how representation learning approaches for audio classification outperform traditional handcrafted feature based approaches.
Moreover, we compared the effect of different design parameters of CNNs in audio classification. 
Text-based approaches seem to outperform other modalities, and benefit from the semantic enrichment of texts via entity linking. %, helping reduce the gap between audio- and text-based approaches.
% However, compared with previous research in the field, the gap between audio and text-based approaches has been reduced. 
% This has been possible thanks to the introduction of deep learning approaches. 
While the image-based classification yielded the lowest performance, it helped to improve the results when combined with other modalities.
Multimodal approaches appear to outperform single modality approaches, and the aggregation of the three modalities achieved the best results.
Furtheremore, the dimensionality reduction of target labels led to better results, not only in terms of accuracy, but also in terms of catalog coverage.

This paper is an initial attempt to study the multi-label classification problem of music genres from different perspectives and using different data modalities. In addition, the release of the \emph{MuMu} dataset opens up a number of unexplored research possibilities. In the near future we aim to modify the ResNet to be able to learn latent factors from images as we did in other modalities and apply the same multimodal approach to other MIR tasks.

\section{Acknowledgments}
This work was partially funded by the Spanish Ministry of Economy and Competitiveness under the Maria de Maeztu Units of Excellence Programme (MDM-2015-0502).
The Tesla K40 used for this research was donated by the NVIDIA Corporation.

% For bibtex users:
\bibliography{ISMIRtemplate}

% For non bibtex users:
%\begin{thebibliography}{citations}
%
%\bibitem {Author:00}
%E. Author.
%``The Title of the Conference Paper,''
%{\it Proceedings of the International Symposium
%on Music Information Retrieval}, pp.~000--111, 2000.
%
%\bibitem{Someone:10}
%A. Someone, B. Someone, and C. Someone.
%``The Title of the Journal Paper,''
%{\it Journal of New Music Research},
%Vol.~A, No.~B, pp.~111--222, 2010.
%
%\bibitem{Someone:04} X. Someone and Y. Someone. {\it Title of the Book},
%    Editorial Acme, Porto, 2012.
%
%\end{thebibliography}

\end{document}